\documentclass[english,aps,twocolumn]{revtex4}
\usepackage{epsfig}
\usepackage{graphicx}
\usepackage{amsmath}
\usepackage{color}

\makeatletter
\@ifundefined{textcolor}{}
{%
 \definecolor{BLACK}{gray}{0}
 \definecolor{WHITE}{gray}{1}
 \definecolor{RED}{rgb}{1,0,0}
 \definecolor{GREEN}{rgb}{0,1,0}
 \definecolor{BLUE}{rgb}{0,0,1}
 \definecolor{CYAN}{cmyk}{1,0,0,0}
 \definecolor{MAGENTA}{cmyk}{0,1,0,0}
 \definecolor{YELLOW}{cmyk}{0,0,1,0}
 }

\makeatother

\usepackage{babel}

\begin{document}

%\preprint{This line only printed with preprint option}

\title{Maris polarization in neutron-rich nuclei}

\author{Shubhchintak}
\affiliation{Department of Physics, Texas A\& M University-Commerce, Commerce, TX 75924, USA}
\author{C.A. Bertulani}
\affiliation{Department of Physics and Astronomy, Texas A\&M University-Commerce, Commerce, TX 75429-3011, USA}
\affiliation{Institut f\"ur Kernphysik, Technische Universit\"at Darmstadt, D-64289 Darmstadt, Germany}
\author{T. Aumann}
\affiliation{Institut f\"ur Kernphysik, Technische Universit\"at Darmstadt, D-64289 Darmstadt, Germany}
\affiliation{GSI Helmholtzzentrum für Schwerionenforschung, Planckstr. 1, 64291 Darmstadt, Germany}
\begin{abstract}
We present a theoretical study of the Maris polarization effect and its application in quasi-free reactions to assess information on the  structure of exotic nuclei. We discuss the uncertainties in the calculations of triple differential cross sections and of analyzing powers due the choices of various nucleon-nucleon interactions the optical potentials and limitations of the method.  Our calculations explore a large number of choices for the nucleon-nucleon (NN) interactions and the optical potential for nucleon-nucleus scattering. Our study implies that polarization variables in (p,2p) reactions in inverse kinematics can be an effective probe of single-particle structure of nuclei in radioactive-beam facilities.
\end{abstract}

\maketitle

Elastic differential cross sections of polarized protons incident on nuclear targets display an interference pattern due to the scattering by the near and the far side of the nucleus. A crucial part of this interference pattern is due to the sign change of the angular momentum in the ${\bf S}\cdot{\bf L}$ spin-orbit part of the optical potential (see, e.g., Ref. \cite{BD04}).   Other types of direct collisions using polarized protons are also influenced by the sign of the spin-orbit part of the optical potential. With the availability of high-energy radioactive beams, quasifree (p,2p) and (p,pn) reactions in inverse kinematics have again become an  experimental tool of choice to study nuclear spectroscopy. Newly developed detectors have allowed efficient experiments using inverse kinematics with hydrogen targets and opened new  possibilities to investigate the single-particle structure, nucleon-nucleon correlations in the nuclear matter, and other important nuclear properties  as the neutron-to-proton ratio of secondary beam projectiles increases. These new developments are possible due to the detection of all outgoing particles, providing kinematically complete measurements of the reactions being carried out at the GSI/Germany, RIKEN/Japan, and other nuclear-physics facilities worldwide \cite{Au07,Kob08,ABR13,Pan16}.  So far, the experiments have focused on the reliability of quasi-free scattering using inverse kinematics as a technique to study the shell-evolution in neutron-rich nuclei, but detailed studies such as the quenching of spectroscopy factors and single-particle properties of neutron-rich nuclei have also been reported recently \cite{Atar17}.  Concomitantly, theoretical interest on (p,2p) reactions is again on the rise \cite{ABR13,Oga15,Mor15,Cra16}.
\begin{figure}[t]
\begin{center}
\includegraphics[
width=3.in]{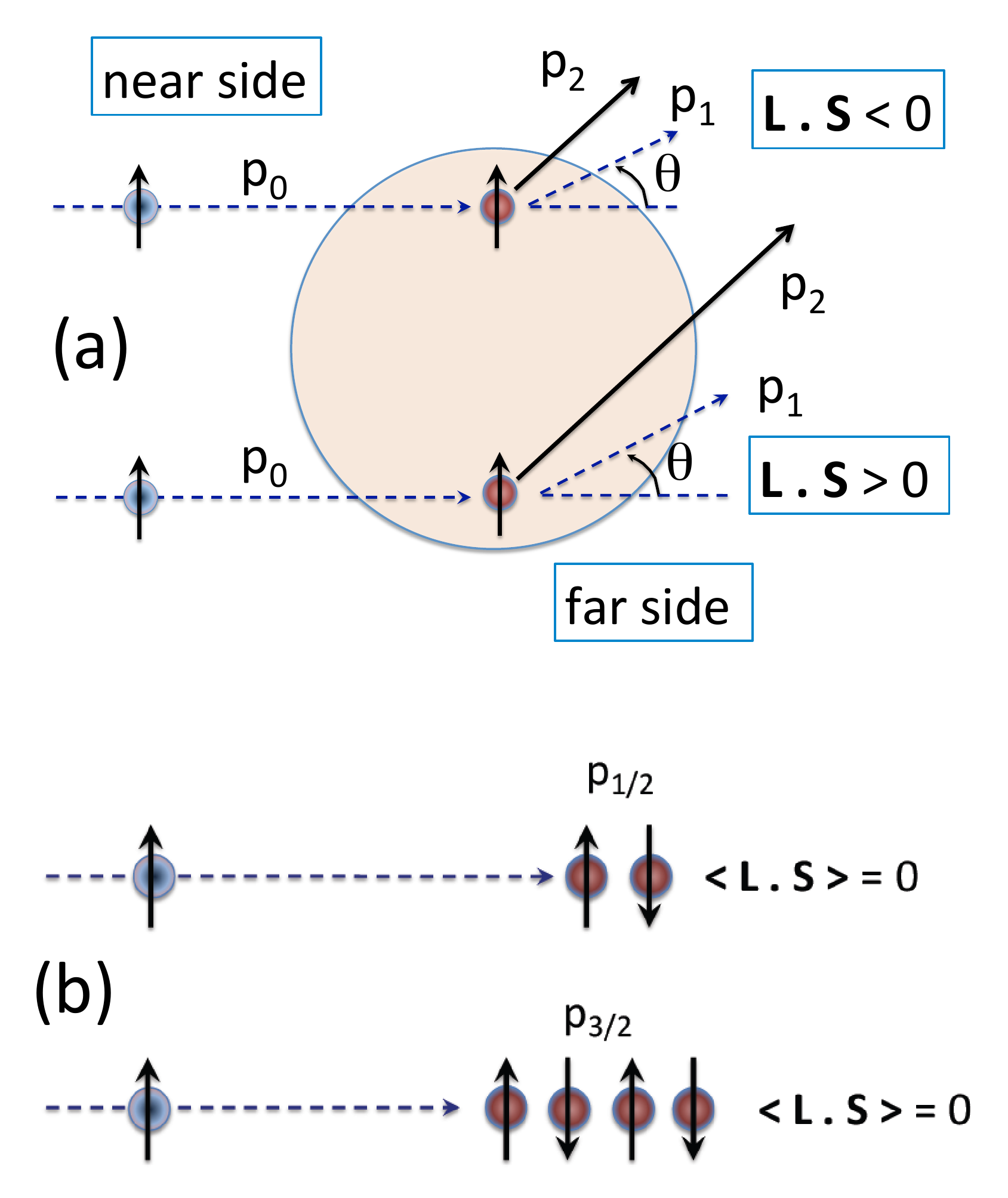}
\end{center}
\caption{(Color online) Top panel (a):  a spin-up proton knocks out a spin-up nucleon (proton or neutron). The proton scattering off the near and the far side leads to opposite signs of the spin-orbit part of the optical potential (OP) as well as to shorter  and longer paths within the nucleus. Bottom panel (b): the collisions within a closed subshell with a spin-independent NN-interaction do not effectively change of the initial proton polarization. A net depolarization of the incident proton occurs with a spin-dependent NN-interaction. This depolarization effect  increases with the number of nucleons in the closed subshell. The final proton polarization will be thus sensitive to the combined effects of the interference between the near and far side scattering caused by the absorption and the spin-orbit parts of the OP, and by the number of nucleons in the subshell.}
\label{marisf}
\end{figure}

In this Letter, we explore the details of the ``Maris effect" \cite{Mar59,Jac76,Mar79,Krei95} systematically in dependence of the neutron-to-proton asymmetry. We show that the effective polarization of knocked out  protons increase steadily with the neutron number. The Maris effect on the spin orientation of the ejected nucleon is caused by the action of the spin-orbit  and absorption parts of the optical potential combined with the distinct occupations in the single-particle  $j_>=l+1/2$ and $j_<=l-1/2$ orbitals \cite{Kel06}. Next, we mention the spin variables of the incident proton, although the same argumentation applies to the knocked-out nucleon. In fact, the Maris polarization effect was proposed as a measure of the polarization of the ejected nucleon. Suppose that the primary spin-up polarized proton is detected at an angle $\theta$, as depicted in Figure \ref{marisf}.  Protons hitting initially polarized spin-up nucleons in a $j$-orbital  with their incoming momenta directed toward the near side, correspond to  ${\bf L}\cdot{\bf S} <0$ and to ${\bf L}\cdot{\bf S} >0$ if the protons are directed to the far side. Because of their smaller path within the nucleus, for collisions happening at the near side the protons will undergo less attenuation than those involved in collisions at the far side. Therefore their initial polarization is modified less than if they were knocked out from the far side. The optical potential dependence on  the ${\bf L}\cdot{\bf S} $ spin-orbit term combined with absorption will thus impact on the polarization changes from near and far side scattering (part (a) of Figure \ref{marisf}). 

The polarization of the incoming proton does not change when the collisions are summed over all nucleons removed from a closed subshell if the momentum distributions of nucleons within the subshell are identical and if the nucleon-nucleon (NN) interaction is spin-independent (part (b) of Figure \ref{marisf}).  However, the NN-interaction has a known spin-dependence for (spin-up)-(spin-up) and (spin-down)-(spin-up) cross sections for the triplet and singlet scattering. Hence, one should expect a change in  the proton polarization due to the subshell occupancy and its effect will be larger if more nucleons occupy that subshell, i.e., twice as large for $p_{3/2}$ than for $p_{1/2}$ subshells. The combination of absorption, the spin-orbit part of the optical potential, and the spin-dependence of the NN-interaction leads to the Maris polarization effect, most evident in the observation of the analyzing power of the scattered protons, 
\begin{equation}
A_y={{d\sigma(\uparrow)-d\sigma(\downarrow)}\over {d\sigma(\uparrow)+d\sigma(\downarrow)}}.
\end{equation} 
Observing $A_y$ requires the detection of the knocked out nucleon by incoming polarized protons with opposite polarizations. It is also expected that the Maris effect is of opposite sign for the $1p_{1/2}$ compared to the  $1p_{3/2}$ orbital. For more details on the Maris polarization effect, and its applications to nuclear spectroscopy, see, e.g., Refs. \cite{Mar59,Jac76,Mar79,Krei95,Kel06}

The Maris polarization effect is a well established experimental tool, e.g. in (p,2p) reaction studies of nuclear medium effects on the NN-interaction   \cite{Mar59,JM73,Jac76,CR77,Ro78,Chan79,Mar79,Nad81,CR83,YNM83,Krei95,Kel06}.  It has also been employed to investigate medium modifications of the nucleon and meson masses and  the meson-nucleon coupling constants in the nuclear medium, motivated by  strong relativistic nuclear fields, deconfinement of quarks, and also  partial chiral symmetry restoration \cite{HI86,Stok93,Mikl13,BR91,Furn92,Hat92,Hat97,SW86}.   It is worthwhile noticing  that there are various distinct spin-orbit interactions  involved in the Maris effect: (a) the spin-orbit part of the optical potential for the nucleon-nucleus scattering, (b) the spin-orbit  interaction responsible for the $j_<$ and $j_>$ occupancy of the knocked out nucleon orbital, and to a lesser extent, (c) the spin-orbit part of the NN-interaction. 

\begin{figure}[t]
\begin{center}
\includegraphics[
width=3.in]{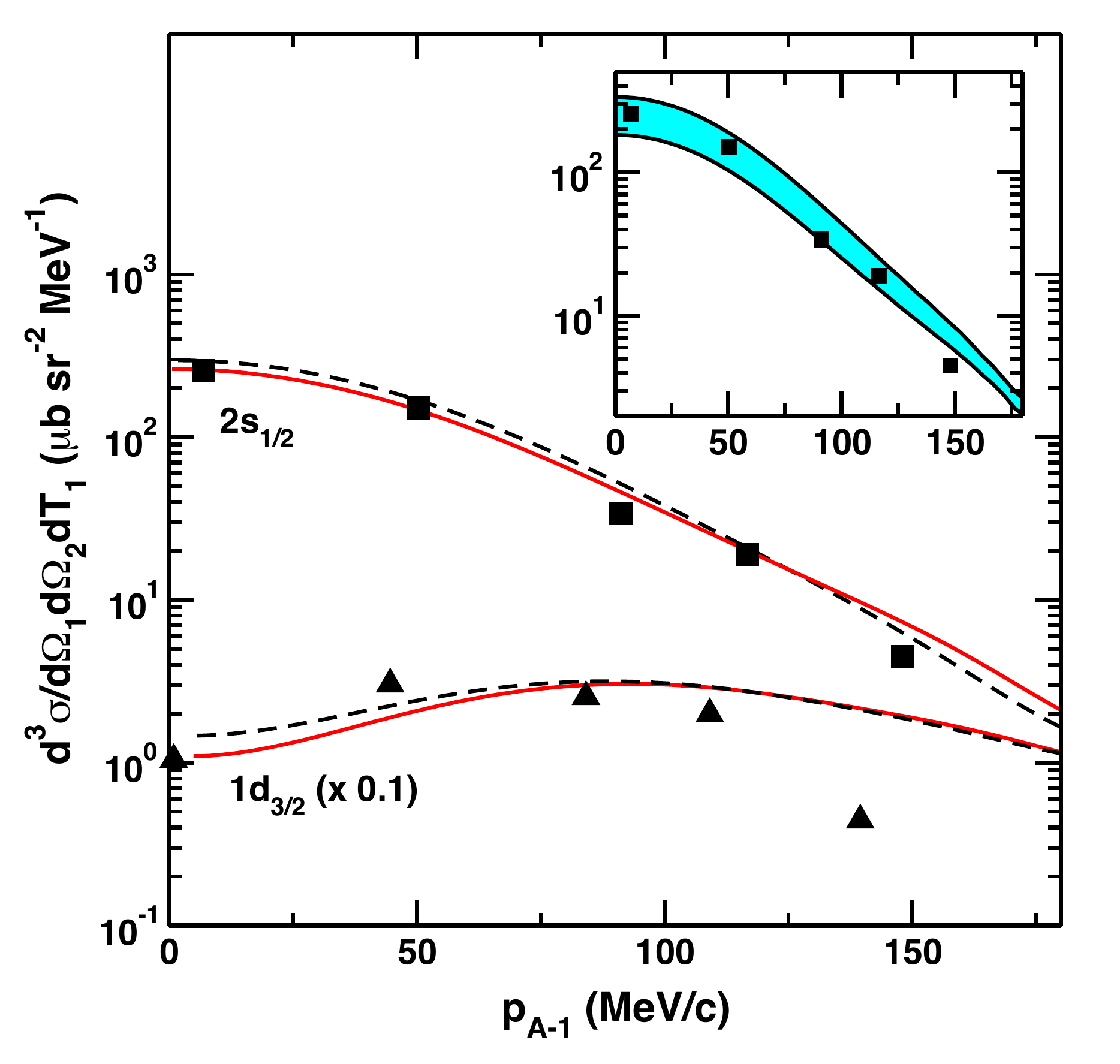}
\end{center}
\caption{(Color online) Cross sections for $^{40}$Ca(p,2p)$^{39}$K with incident proton energy $E_p= 148$ MeV, as a function of the recoil momentum, $p_{A-1}$ of the residual nucleus. The proton knockout is assumed to be either from the $1d_{3/2}$ or from the $2s_{1/2}$ orbital in $^{40}$Ca. The cross sections are integrated over the energy of  the knocked-out proton and given in units of $\mu$b sr$^{-2}$ MeV$^{-1}$. The experimental data are taken from  Ref. \cite{Ro78}. The dashed (solid) lines include (do not include) the spin-orbit part of the optical potential. The inset panel shows that larger uncertainties arise, e.g., for the $1s_{1/2}$ state, with the inclusion of various nucleon-nucleon (NN) interactions. The shaded area contains a broad range of results obtained with different NN-interactions taken from Refs. \cite{YNM83,FL85,NL88,Ke89,Ray90,Stok93,Amo00}.}
\label{fig1}
\end{figure}
The triple differential cross section for quasifree scattering in the Distorted Wave Impulse Approximation (DWIA) is given by \cite{JM73}  
\begin{eqnarray}
{d^3\sigma \over  d\Omega_1 d\Omega_2dT_1} &=&C^2S\cdot K_F \nonumber \\
&\times& \left| \left< \chi_{\sigma_2 {\bf p}_{2}}^{(-)} \chi_{\sigma_1 {\bf p}_{1}}^{(-)}          
\left| \tau_{pN}\right| \chi_{\sigma_0 {\bf p}_0}^{(+)}\psi_{jlm}\right>\right|^2 , \label{tripleX}
\end{eqnarray}
where $K_F$ is a kinematic factor,  ${\bf p}_0$ (${\bf p}_1$) denotes the momentum of the incoming (outgoing) proton, ${\bf p}_2$ the momentum of the knocked-out nucleon, and $T_{2}$ its  energy. $C^2S$ is the spectroscopic factor associated with the single-particle properties of the removed nucleon and $\psi_{jlm}$ is its wavefunction,  labelled by the $jlm$  quantum numbers. The DWIA matrix element includes the scattering waves $\chi_{\sigma {\bf p}}$ for the incoming and outgoing nucleons, with information on their spins and momenta, ($\sigma {\bf k}$), as well and the t-matrix for the nucleon-nucleon scattering. To first-order this t-matrix is directly proportional to the free NN scattering t-matrix, $\tau_{pN}$.  For unpolarized protons, Eq. \eqref{tripleX} is averaged over initial  and summed over final spin orientations. This formalism has been used previously  and a good description of experimental data has been obtained with a proper choice of the optical potential and of the NN-interaction (see, e.g., Refs. \cite{CR77,CR83}). In Ref. \cite{ABR13} it was shown that momentum distributions of the residual nuclei obtained in quasi-free scattering are well described using the eikonal approximation for the scattering waves $\chi_{{\bf p}_i}$ entering Eq. \eqref{tripleX}. The method, appropriate for high-energy collisions, allows to easily include relativistic and medium effects and a connection with partial waves can be done for large angular momenta with $L=pb$, where $p$ is the incident momentum and $b$ the impact parameter. Here, we adopt the DWIA and the partial wave expansion method described in various publications, e.g., Refs. \cite{Mar59,Jac76,Mar79,HI86,Krei95,Hat97,JM73,CR77,CR83}. 

The inputs for the calculations following Eq. \eqref{tripleX} are (a) the optical potential for nucleus-nucleus scattering, (b) the NN-interaction, and (c) the ejected nucleon wavefunction $\psi_{jlm}$.  For simplicity, the single-particle energies and wavefunctions $\psi_{jlm}$ of the ejected nucleon are calculated with a  global Woods-Saxon potential model in the form $V(r)=[V_0  + (0.72\ {\rm fm}^2)V_{SO}/(ar)]f(r)$, $f(r)=\left\{1+\exp[(r-R)/a]\right\}^{-1}$,  $V_0=[-57.8\pm 33(N-Z)/A]$ MeV with +($-$) sign for neutrons (protons), and   $V_{SO}=[-22\pm 14(N-Z)/A]$ MeV. We use $a=0.65$ fm and $R=1.2A^{1/3}$ fm.

In Figure \ref{fig1}  we show the calculated cross sections for $^{40}$Ca(p,2p)$^{39}$K and incident proton energy $E_p= 148$ MeV, as a function of the recoil momentum, $p_{A-1}$ of the residual nucleus. The proton knockout is assumed to be from the $1d_{3/2}$ and $2s_{1/2}$ orbitals in $^{40}$Ca.   The cross sections are integrated over the energy of the knocked-out proton and are given in units of $\mu$b sr$^{-2}$ MeV$^{-1}$. The optical potential of Ref. \cite{Nad81} and the NN-interaction of Ref. \cite{Stok93} were employed. The experimental data are taken from  Ref. \cite{Ro78}. The dashed (solid) lines include (do not include) the spin-orbit  part of the optical potential.  In agreement with the conclusions of Refs. \cite{Ro78,Chan79}, we find that  the spin-orbit part of the optical potential plays a small role in the description of the triple-differential cross sections for unpolarized protons.  

The inset panel in Figure \ref{fig1} shows a comparison of our calculations with the experimental data of Ref. \cite{Ro78} for the $1s_{1/2}$ state as various NN-interactions are used.  The shaded area includes results for seven NN-interactions taken from Refs. \cite{YNM83,FL85,NL88,Ke89,Ray90,Stok93,Amo00}. We have observed that the choice of the NN-interaction has a greater impact on the results for unpolarized protons than the strength of the spin-orbit part of the optical potential. The same conclusion applies for the proton removal from the 1$d_{3/2}$ orbital.  Similarly, different choices for the other parts of the optical potential adopted also yield  a broad range of results. We will discuss this problem again in the context of the Maris effect. 

\begin{figure}[t]
\begin{center}
\includegraphics[
width=3.in]{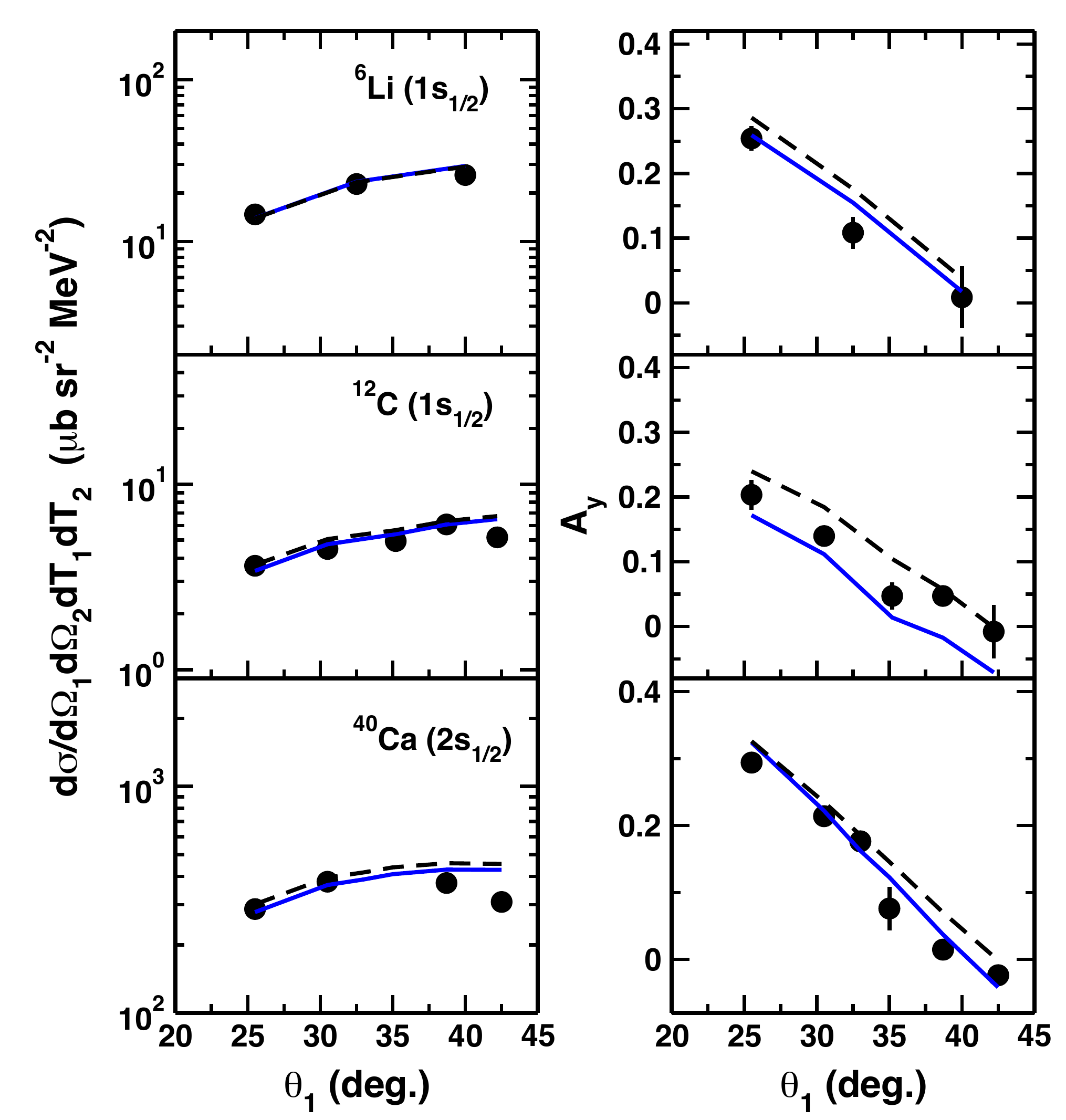}
\end{center}
\caption{(Color online) Triple-differential cross sections (left) and analyzing powers (right) for (p,2p) reactions with $^6$Li, $^{12}$C and $^{40}$Ca at proton energy of 392 MeV. The solid lines include the spin-orbit part of the optical potential, the dashed lines display the results without the spin-orbit part. The data are taken from Ref. \cite{Hat97}.}
\label{fig2}
\end{figure}

\begin{figure}[t]
\begin{center}
\includegraphics[
width=2.9in]{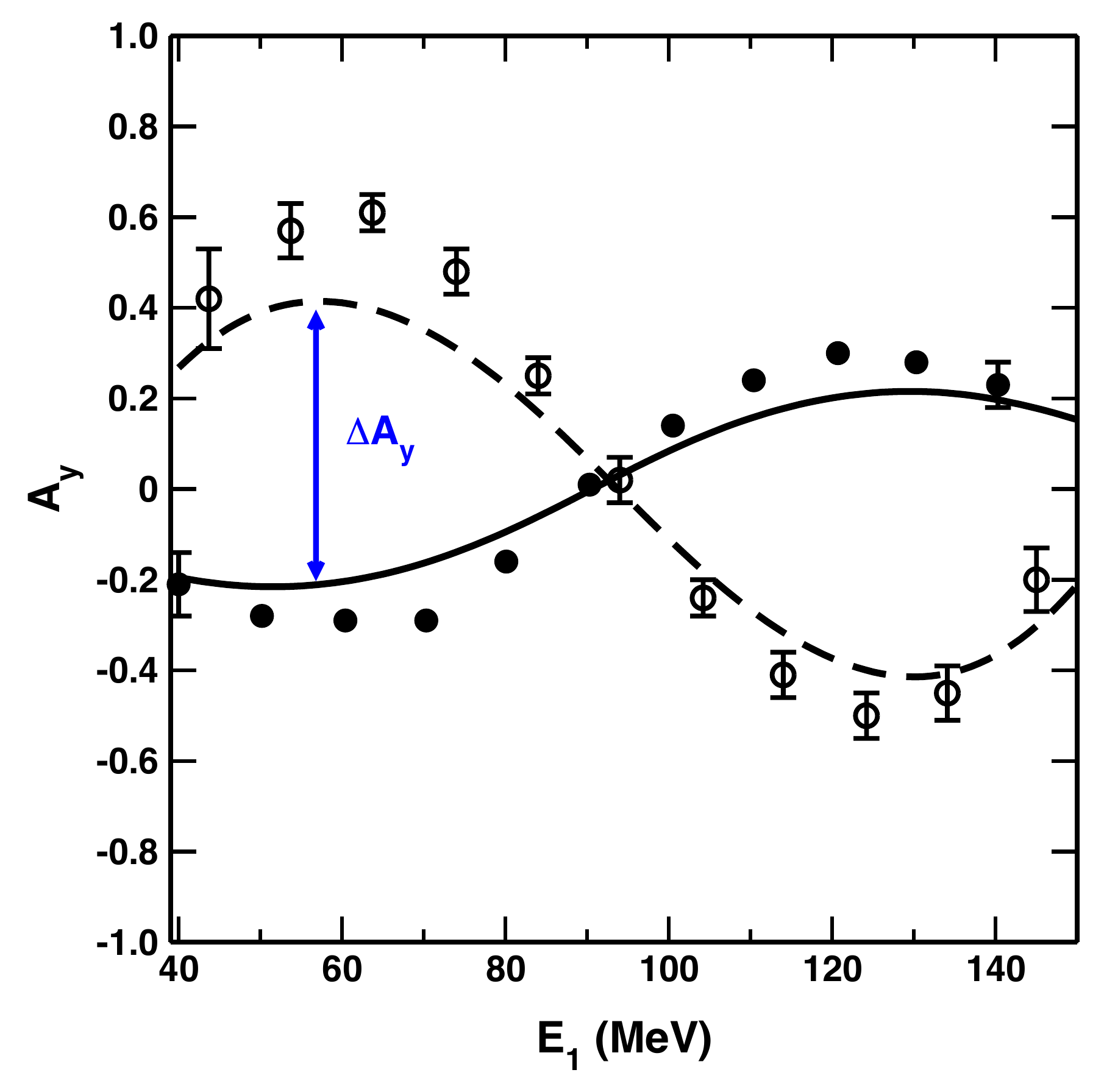}
\end{center}
\caption{(Color online) Analyzing powers for proton knockout from the $1p_{3/2}$ and $1p_{1/2}$ states  in the $^{16}$O(p,2p) reaction at 200 MeV as a function of the kinetic energy of the ejected proton.  One proton is measured at 30$^\circ$ and the other at $-30^\circ$. The open circles  are  for the $1p_{1/2}$ and the solid circles  for the $1p_{3/2}$ orbital. }
\label{fig3}
\end{figure}

In Figure \ref{fig2} we show our calculations for the triple-differential cross sections (left) and analyzing powers (right) in (p,2p) reactions with $^6$Li, $^{12}$C and $^{40}$Ca at incident proton energy of 392 MeV. The data are taken from Ref. \cite{Hat97}. To achieve a reasonable agreement with the experimental data, we use the NN interaction from Ref. \cite{Hor85} and the Dirac phenomenological optical potential  from Ref. \cite{Coo93}. The solid lines include the spin-orbit part of the optical potential and the calculations have been normalized to the data  for  $d^3\sigma / d\Omega_1 d\Omega_2 dT_1 $. Due to the nature of the data analysis \cite{Hat97}, we do not try to identify  the normalization values as spectroscopic factors. The dashed lines display our calculations without the spin-orbit part of the optical potential. Protons removed from the s-shell are chosen  because the interpretation is rather simple as the Maris polarization should be null (for the knocked out nucleon, ${\bf S}=0$ and thus ${\bf L}\cdot{\bf S} =0$), although the knocked out proton can still acquire a non-zero angular momentum with respect to the (A-1) residue after the collision due to Final State Interactions (FSI). In fact,  we observe that  the spin-orbit part of the optical potential still plays a small but non-negligible role in our results. 

\begin{figure}[t]
\begin{center}
\includegraphics[
width=2.9in]{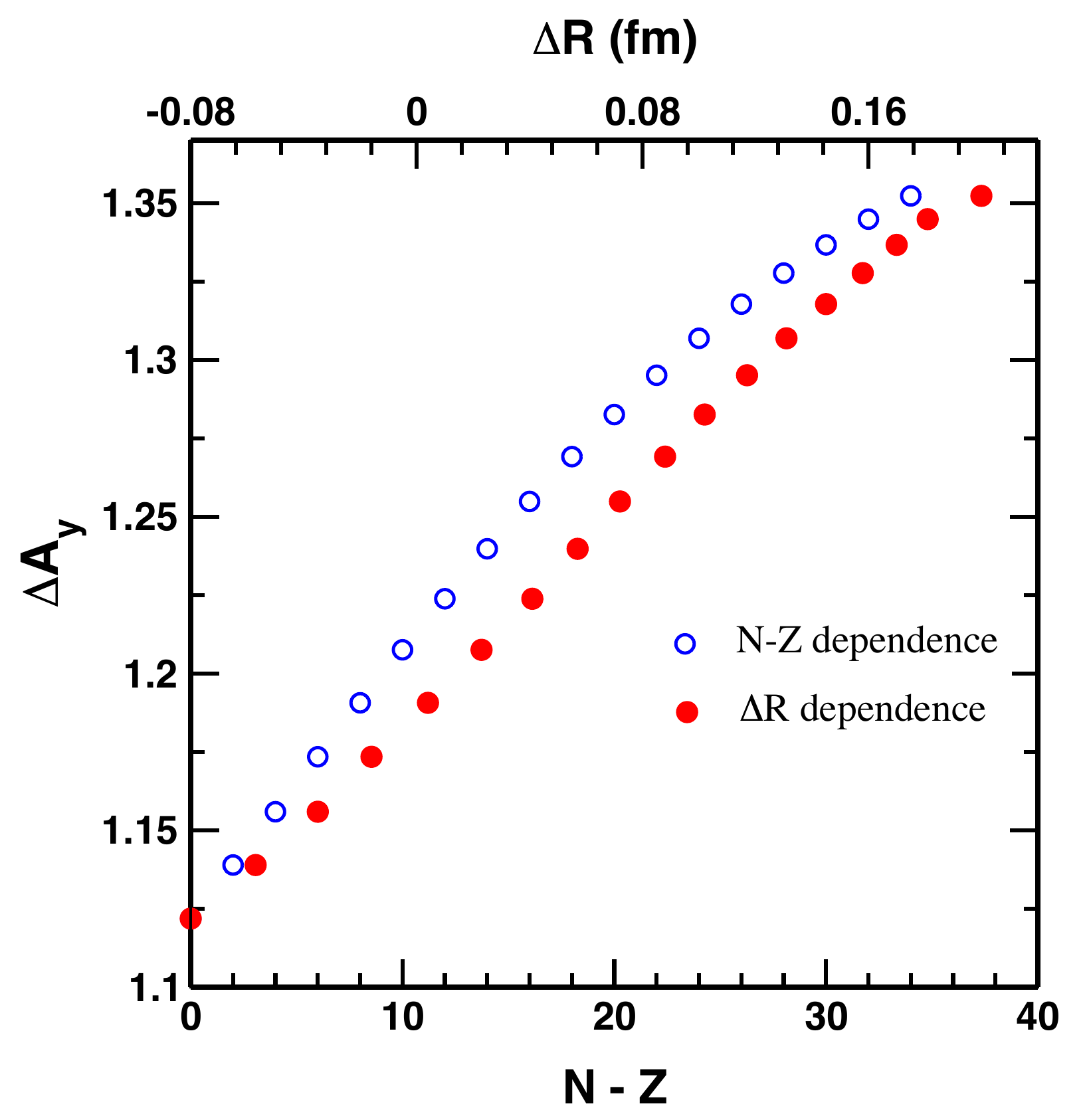}
\end{center}
\caption{(Color online)  Difference in the polarization maximum and minimum (see Figure \ref{fig3}) for the $2p_{1/2}$ and $2p_{3/2}$ subshells in tin isotopes for (p,2p) reactions at $E_p=200$ MeV.  The solid  circles show the calculated $\Delta A_y$ from Eq. \eqref{DAy} as a function of the neutron skin in the nuclei (using upper axis scale). The open circles show the calculated $\Delta A_y$   as a function of the neutron excess (using lower axis scale). We assume that protons are detected at $\theta = 35^\circ$ and  $\theta = -35^\circ$, respectively.}
\label{fig4}
\end{figure}

As suggested in Ref. \cite{Mar79}, the Maris polarization effect should be manifest in measurements of $A_y$, i.e., it should be visible in analyzing power data, specially for protons removed from p-orbitals. This is best seen if $A_y$ is displayed for fixed angles of the outgoing protons while scanning the energy of the ejected proton, as seen in Figure \ref{fig3}.  The data are from Ref. \cite{Kit80}.  One proton is measured at 30$^\circ$ and the other at $-30^\circ$. The open circles are data for protons removed from the $1p_{1/2}$  and the solid ones from the $1p_{3/2}$ orbital.   In our calculations, shown by dashed and solid lines, we have employed the same NN-interaction and optical potential model as in the calculations presented in Fig. \ref{fig1}. 

\begin{figure}[t]
\begin{center}
\includegraphics[
width=2.9in]{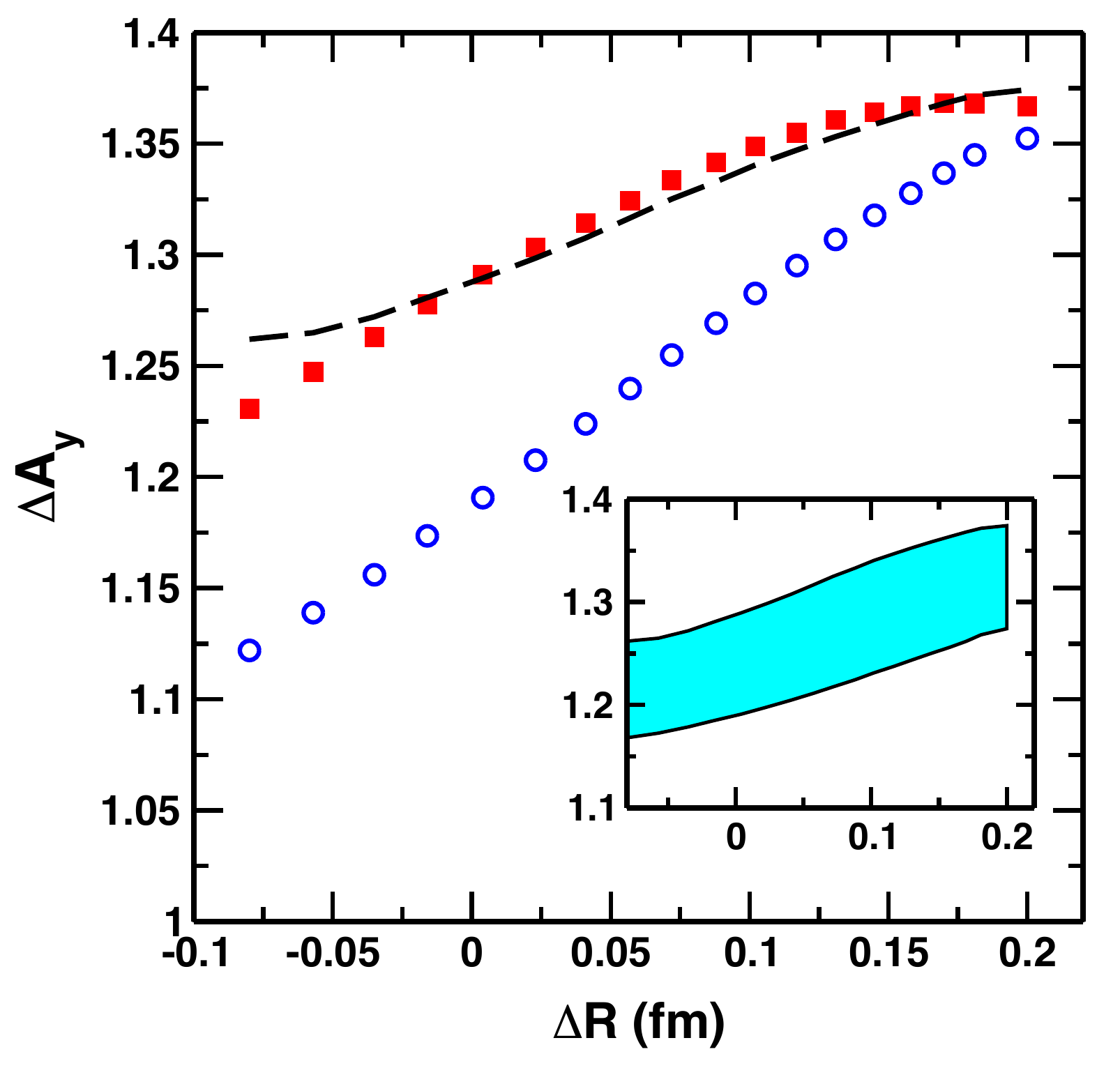}
\end{center}
\caption{(Color online) Difference in the polarization maximum and minimum for the $2p_{1/2}$ and $2p_{3/2}$ subshells in tin isotopes for (p,2p) reactions at 200 MeV as a function of the neutron skin. The open circles are calculated using HFB densities, whereas the squares use sharp-cutoff densities. In this case the normalized proton and neutron sharp-cutoff densities are assumed to have the same neutron skin $\Delta R$ as those obtained with the HFB densities.  The inset panel  shows  calculations for the same case as those performed for the dashed curve in the larger panel but with various optical potentials \cite{Nad81,Lej76,Jack79,FL85,Ham90,Coo93,Kob85,Kon03}. We assume that protons are detected at $\theta = 35^\circ$ and  $\theta = -35^\circ$, respectively. The  dashed curve shows calculations using a single  sharp-cutoff density adding up the proton and neutron densities and a single nuclear radius equal to $R+\Delta R/2$, with $\Delta R$ calculated from the HFB densities. }
\label{fig5}
\end{figure}

As the number of neutrons increases in an isotopic chain, the nuclei should develop a larger neutron skin. The charge distribution in stable nuclei is well determined via electron scattering experiments but similar experiments on  unstable nuclei are very difficult, still far from being fully viable \cite{Ber07,Ant11,SS17,Kar17}.  The determination of the neutron skin in a nucleus requires separate measurements of the matter density. Efforts in this direction involve the measurement of interaction cross sections \cite{Tan85}, total neutron-removal cross sections \cite{Aum17}, parity violation in weak interaction with electron scattering \cite{Abra12}, Coulomb dissociation \cite{Ros13}, antiprotonic atoms \cite{Trz01}, dipole polarization in (p,p') scattering \cite{Tam11}, etc. The analyzing power, being a ratio of cross sections, factors out some of the uncertainties associated in the calculations.  Moreover, because the spin-orbit part of the optical potential is peaked at the nuclear surface, the Maris effect is more sensitive to the surface region of the nucleus than the cross sections for unpolarized protons. Since the ejected nucleon spin will be depolarized more and more by the absorption effect when the nuclear size and neutron skin increases, a dependence of the Maris polarization with the neutron-skin thickness could be expected.

Based on the arguments above, we consider the Maris polarization effect in neutron-rich nuclei and its dependence on the neutron number along a typical isotopic chain, e.g., for tin isotopes. Our calculations are not intended to be accurate, but to use the state-of-the-art theoretical knowledge one has on nuclear densities to explore the evolution of the Maris effect with the variation of the neutron skin. Most global optical potentials for proton-nucleus scattering reflect nuclear sizes and their dependence on the total number of nucleons, being insensitive to the build-up of a neutron skin in the nuclei. In order to study the role of the nuclear density and its neutron skin, we construct an optical potential from a folding model of the nuclear density with an effective nucleon-nucleon interaction. We chose the well-known Franey-Love interaction \cite{FL85}.   For the nuclear densities we adopt two models: (a)  densities calculated with the Hartree-Fock-Bolgoliubov (HFB) method and with the BSk2 Skyrme interaction as described in Ref. \cite{Sam02}, and (b) with constant densities up to a sharp-cutoff radius. The microscopic HFB calculations are used to estimate the neutron skin of the nuclei along the isotopic chain. The neutron skin, defined as $\Delta R=\sqrt{< r_n^2 >} - \sqrt{< r_p^2 >}$ is extracted from the HFB calculations and used in part (b) of the prescription above to generate (properly normalized) proton and neutron sharp-cutoff densities.  

We quantify the magnitude of the Maris polarization in terms of the  difference between the first maximum of the $2p_{1/2}$ and the first minimum of the $2p_{3/2}$ orbital, denoted by 
\begin{equation}
\Delta A_y= (A_y^{p_{1/2}})_{max} - (A_y^{p_{3/2}})_{min}. \label{DAy}
\end{equation} 
The choice of the $2p_{1/2}$  and $2p_{3/2}$ orbitals to explore the effect of neutron excess is arbitrary.  But it is worthwhile mentioning that the single-particle 2p orbitals in tin isotopes are probably highly fragmented. This would have to be taken into consideration in future experiments. The single-particle wavefunctions $\psi_{jlm}$ for these orbitals could be extracted from the HFB calculations, but for convenience we adopt the global Woods-Saxon potential defined previously to calculate the bound states along the tin isotopic chain. All 2p orbitals in tin are bound within this approximation.

In Figure \ref{fig4} we plot $\Delta A_y$ for (p,2p) reactions at $E_p=200$ MeV  with the densities defined in (a) and (b) discussed above. We assume that the two protons are detected at $\theta = 35^\circ$ and $\theta = -35^\circ$, respectively. Using the lower scale, the graph shows the dependence of the observable in Eq. \eqref{DAy} as a function of the neutron excess (open circles), while the upper scale shows the same quantity as a function of the neutron skin (closed circles).  These results imply that the increasing neutron number in an isotope leads to a larger magnitude of the Maris polarization effect. The effective polarization increases by more than 30\% along the tin isotopic chain. The dependence with the neutron skin is almost linear, although deviations from the linear proportionality appears for large neutron excess. Since the proton density radius is nearly constant along the isotopic chain, as estimated with the HFB calculations, the steady increase of $\Delta A_y$ is a clue for the build-up of neutrons at the nuclear surface. 

In Figure \ref{fig5} we show a comparison between the calculations displayed in Figure \ref{fig4} for Eq. \eqref{DAy} (open circles) with those using sharp-cutoff densities, displayed as red squares in the figure. In this case the normalized proton and neutron sharp-cutoff densities are assumed to have the same neutron skin $\Delta R$ as those obtained with the HFB densities. There are appreciable differences between the two calculations reflecting the fact that the quantity defined in Eq. \eqref{DAy} is also sensitive to the details of the densities such as their diffuseness. 

In Figure \ref{fig5} we also show a  dashed curve calculated with a single  sharp-cutoff density adding up the proton and neutron densities and a single nuclear radius equal to $R+\Delta R/2$, with $\Delta R$ calculated from the HFB densities. Despite small deviations from the previous result displayed as red squares in the figure, the $\Delta A_y$ increase along the isotopic chain for the dashed-line is also representative of the increase of the nuclear radius, irrespective if the nuclear densities display a neutron skin or not. 

The inset panel in Figure \ref{fig5}  shows  calculations for the same case as that performed for the dashed curve in the larger panel but now, for the inset, we adopt a plethora of optical potentials \cite{Nad81,Lej76,Jack79,FL85,Ham90,Coo93,Kob85,Kon03}. We observe a strong dependence of $\Delta A_y$ on the optical potential adopted, as expected. Nonetheless, $\Delta A_y$ still displays an increase with the neutron number in the isotope. We have also noticed that a similar result and conclusion is obtained for its dependence of $\Delta A_y$ on various NN interactions, i.e., $\Delta A_y$ is also strongly dependent on the NN-interaction used. Therefore, using $\Delta A_y$ as a probe of the nuclear size or the neutron skin in nuclei invokes a complementary study of other observables to determine the optical potential parameters as well as the adequate NN-interaction to be used in the theory.

In conclusion, the Maris polarization effect is well known as a tool to investigate single-particle properties in nuclei. It has not been widely explored yet to study the evolution of nuclear properties in neutron-rich isotopes. Its sensitivity to the shell occupancy of orbitals with the same angular momentum allows for new applications in experimental studies carried out with secondary radioactive beams.  Because experiments can now be carried out with a much larger precision than in the past, new techniques are increasingly being introduced to extend our knowledge of the nuclear physics of neutron-rich nuclei. We demonstrate that the magnitude of the Maris polarization effect increases with the neutron excess. However, the increasing magnitude of the effect cannot be related in a straightforward manner to the development of the  neutron-skin thickness in  neutron-rich nuclei, but rather depends as well on the size of the nucleus and also on the diffuseness of the densities at the surface. The slope of the dependence of the calculated analyzing power with the neutron excess does not vary substantially, neither with the selection of the NN interaction or with the optical potential. But, in contrast, its absolute magnitude does show a strong dependence on the choice of these two interactions.

This work was supported in part by the U.S. DOE grant DE- FG02-08ER41533 and the U.S. NSF Grant No. 1415656. We thank HIC for FAIR for supporting visits (C.A.B.) to the TU-Darmstadt.

\end{document}